\documentclass[12pt]{article}

\usepackage{amsmath,amssymb,amsthm,epsf,epsfig}
\usepackage{amsmath,amssymb}

\begin{document}
{\bf \underline{Physica Scripta 79 (2009) 065003}}
\begin{center}
\Large {\bf Coherent state of nonlinear oscillator and it's
revival dynamics}
\end{center}
 \vspace{.2cm}

\begin{center}
{\bf \it B.Midya{\footnote{e-mail : bikash.midya@gmail.com}} , B. Roy{\footnote{e-mail : barnana@isical.ac.in}}\\
Physics and Applied Mathematics Unit \\
Indian Statistical Institute \\
Kolkata - 700 108, India.} \\

\vspace*{0.2cm}

{\bf \it A.Biswas{\footnote {e-mail : atreyee11@gmail.com}}\\
Department of Natural Sciences and Humanities\\
West Bengal University of Technology\\
Salt Lake City, Kolkata-700064, India}\\
\end{center}

\vspace{1 cm}

\begin{center} {\large {\bf {Abstract}}} \end{center}
The coherent state of the nonlinear oscillator having nonlinear spectrum, is constructed using the Gazeau-Klauder formalism. The weighting distribution, the Mandel parameter are studied. The details of the revival structure arising from different time scales underlying the quadratic energy spectrum are investigated by the phase analysis of the autocorrelation function.

\section{Introduction}
The nonlinear differential equation
\begin{equation}
(1+\lambda x^2)\ddot {x} - (\lambda x)\dot {x}^2 + \alpha^2 x = 0,~~~\lambda > 0
\end{equation}
was studied by Mathews and Lakshmanan in [1,2] as an example of a non-linear oscillator and it was shown that the solution of (1) is
\begin{equation}
x = A sin(\omega t + \phi)\label{sol1}
\end{equation}
with the following additional restriction linking frequency and amplitude
\begin{equation}
\omega^2 = \frac{\alpha^2}{1+\lambda A^2}\label{free}
\end{equation}
It has been shown in [1] that Eqn.(1) is obtainable from the Lagrangian density
\begin{equation}
L = \frac{1}{2}\frac{1}{(1+\lambda x^2)}(\dot {x}^2 - \alpha^2
x^2)\label{lag}
\end{equation}
Recently in a series of papers [3-5] this particular nonlinear system has been generalized to the higher dimensions and various properties of this system have been studied thoroughly. In ref [6], the Schr\"odinger equatiion corresponding to this nonlinear oscillator has been solved exactly as a Sturm-Liouville problem and $\lambda$-dependent eigenvalues and eigenfunctions were obtained for both $\lambda > 0 $ and $\lambda < 0$. It is to be noted that\\
1) this $\lambda$-dependent system can be considered as a
deformation of the standard harmonic oscillator in the sense that
for $\lambda \rightarrow 0$ all the characteristics of the linear
oscillator are recovered.\\
 2) in the Lagrangian (4), the
parameter $\lambda$ is present not only in the potential
$(\frac{x^2}{1+\lambda x^2})$ but also in the kinetic term. So
this nonlinear oscillator may also be considered as a particular
system with position dependent mass $m(x) = \frac{1}{(1+\lambda
x^2)}$.

 Schr\"odinger equation with a
position dependent mass has found applications in the field of
material science and condensed matter physics, such as
semiconductors [7], quantum wells and quantum dots [8,9], 3He
clusters [10], graded alloys and semiconductor heterostructures
[11-19] etc. It has also been found that such equations appear in
very different areas.For example, it has been shown that constant
mass Schr\"odinger equation in curved space and those based on
deformed commutation relations can be interpreted in terms of
position dependent mass [20,21]. This has generated a lot of
interest in this field and during the past few years various
theoretical aspects of position dependent mass Schr\"odinger
equation have been studied widely [22-39].

 On the
other hand coherent states have attracted considerable attention
in the literature [40,41]. Coherent states are generally
constructed by (i) using the displacement operator technique or
defining them as (ii) mininum uncertainty states or (iii)
annihilation operator eigenstates. However, even when such
operators do not exist, different approaches [42-47] have been
utilized to construct coherent states corresponding to different
quantum mechanical potentials [48-74]. Coherent states of systems
possessing nonlinear energy spectra are of particular interest as
their temporal evolution can lead to revival and fractional
revival, leading to Schr\"odinger cat and cat-like states. For
constant mass Schr\"odinger equation, potentials like,
P\"oschl-Teller, Morse and Rosen-Morse lead to nonlinear spectra.
Time evolution of the coherent states for these potentials [75-83]
is a subject of considerable current interest as they can produce
Schr\"odinger cat and cat-like states.

In this paper we shall construct coherent states for the nonlinear
oscillator (1) and study their revival dynamics. The motivation
comes from the fact that the study of the temporal evolution of a
free wave-packet with position dependent mass inside an infinite
well [84] has revealed that revival and partial revivals are not
only different from the constant mass case but also they are very
much dependent on the mass function $m(x)$. So it will be
interesting to study the role of the mass parameter $\lambda$ of
the nonlinear oscillator (1) on the temporal evolution of it's
coherent state.

 The paper is organized as follows. In section 2.
the Gazeau-Klauder coherent state [45,46] for the nonlinear
oscillator is constructed and is shown to satisfy the conditions
such as, continuity of labelling, resolution of unity, temporal
stability and action identity. The revival dynamics of the
coherent state is studied in section 3. In section 4. we summarize
our results.

\section {\bf Gazeau-Klauder coherent state for nonlinear oscillator}

The position dependent mass Schr\"odinger Hamiltonian $H$ for the nonlinear oscillator(1) is given by
\begin{equation}
H = \left[-\frac{\hbar ^2}{2m}\left(1+\lambda x^2 \right)\frac{d^2}{dx^2} - \frac{\hbar ^2}{2m} \lambda x \frac{d}{dx} + \frac{1}{2} g\left(\frac{x^2}{1+\lambda x^2}\right)\right]
\end{equation}
The parameter $\lambda$ may be positive or negative. However for $\lambda>0$ the discrete energy spectrum can be shown to be finite and consequently for completeness property the continuum has to be taken into account. On the other hand for $\lambda<0$ there are only discrete energy states and henceforth we shall consider this choice. Replacing the parameter $g$ in Eqn.(5) by $m\alpha^2+\lambda \hbar \alpha$ as was done in ref[6], the Schr\"odinger equation is given by
\begin{equation}
\left[\frac{-\hbar ^2}{2m}(1+\lambda x^2)\frac{d^2}{dx^2} - \frac{\hbar ^2}{2m} \lambda x \frac{d}{dx} + \frac{1}{2}m\alpha(\alpha + \frac{\hbar \lambda}{m})(\frac{x^2}{1+\lambda x^2})\right]\psi = E\psi
\end{equation}
Under the transformation $ x = \sqrt{\frac{\hbar}{m\alpha}}y$~~;~~$\lambda = \frac{m\alpha}{\hbar}\Lambda$, the above equation transforms to
\begin{equation}
\left[(1+\Lambda y^2)\frac{d^2}{dy^2} + \Lambda y \frac{d}{dy} -(1+\Lambda)\frac{y^2}{(1+\Lambda y^2)} + 2e\right]\psi = 0
\end{equation}
where $E = e(\hbar \alpha)$.
For $\Lambda < 0$, the eigenvalues $e_n$ and the eigenfunctions $\psi_n$ of the above transformed Hamiltonian are given by [6]
\begin{equation}
\psi_n = (1-|\Lambda|y^2)^{\frac{1}{(2|\Lambda|)}}{\cal {H}}_n(y,\Lambda),~~~~e_n = (n+\frac{1}{2}) + \frac{1}{2}n^2 |\Lambda|,~~n=0,1,2 \cdots m \cdots
\end{equation}
where ${\cal {H}}_n(y,\Lambda)$ are called $\Lambda$-deformed Hermite polynomial [6].
So the eigenvalues of the Hamiltonian $H^1 = H - \frac{\hbar \alpha}{2}$ are
\begin{equation}
E_n^1 = E_n - \frac{\hbar\alpha}{2} = \hbar\alpha\left[n+\frac{1}{2}n^2|\Lambda|\right]= \hbar\alpha\frac{n(n+\mu)}{\mu} = \hbar\alpha e_n^1
\end{equation}
where $\mu = \frac{2}{|\Lambda|}$.
The Gazeau-Klauder coherent state [45,46] for this system is given by
\begin{equation}
|J,\gamma> = \frac{1}{N(J)}\sum_n \frac{(J)^{\frac{n}{2}}exp(-i\gamma e_n^1)}{\sqrt{\rho_n}}|n>
\end{equation}
where $\gamma = \alpha t$. The normalisation constant $N(J)$ is given by
\begin{equation}
N(J) = \left[\sum_n \frac{J^n}{\rho_n}\right]^{\frac{1}{2}}~~~,0<J<R = \lim sup_{n \rightarrow +\infty}\rho_n^{\frac{1}{n}}
\end{equation}
where $R$ denotes the radius of convergence and $\rho_n$ denotes the moments of a probability distribution $\rho(x)$
\begin{equation}
\rho_n = \int_0^R x^n \rho(x)dx = \prod_{i=1}^n e_i^1~~;~~\rho_0 = 1
\end{equation}
For the coherent state (10)
\begin{equation}
\rho_n = \prod_{i=1}^n \frac{i(i+\mu)}{\mu} = \frac{\Gamma (n+1)\Gamma (n+1+\mu)}{\mu^n \Gamma(1+\mu)}~~,\rho_0 =1
\end{equation}
so that $R$ is infinite and
\begin{equation}
\rho(J) = \frac{2\mu(J\mu)^{\frac{\mu}{2}}}{\Gamma (1+\mu)}K_{\mu}(2\sqrt{J\mu})
\end{equation}
$K_{\nu}(cx)$ being modified Bessel function [85].\\
Also
\begin{equation}
N(J)^2 = \frac{\Gamma (1+\mu)}{(J\mu)^{\frac{\mu}{2}}}I_{\mu}(2\sqrt{J\mu})
\end{equation}
where $I_{\mu}(cx)$ is the modified Bessel function [85].\\
So the coherent state (10) finally becomes
\begin{equation}
|J,\gamma> = \frac{\sqrt{\Gamma (1+\mu)}}{N(J)}\sum_{n=0}^{\infty} \frac{(J\mu)^{\frac{n}{2}}e^{-i\alpha (n+\frac{n^2}{\mu})t}}{\sqrt{n!\Gamma (n+1+\mu)}}|n>
\end{equation}
Below we shall see that the coherent state (10) satisfies the following conditions:\\
1. Continuity of labelling: From the definition (10) it is obvious that\\
$(J^{\prime} \gamma ^{\prime}) \rightarrow (J,\gamma) \Rightarrow |J^{\prime},\gamma^{\prime}> \rightarrow |J,\gamma>$\\
2.Resolution of unity:
\begin{equation}
\int|J,\gamma><J,\gamma|d\mu(J,\gamma) = \int|J,\gamma><J,\gamma|d\mu(J,\gamma) = \frac{1}{2\pi}\int_{-\pi}^{\pi}d\gamma \int_0^{\infty}k(J)|J,\gamma><J,\gamma|dJ
\end{equation}
where $k(J)$ is defined by
\begin{equation}
\begin{array}{lcl}
k(J) &=& N(J)^2\rho(J) \geq 0,~0\leq J <R\\
     &=& \rho(J)\equiv 0, J>R
\end{array}
\end{equation}
For the coherent state (10)
\begin{equation}
k(J) = 2\mu I_{\mu}(2\sqrt{J\mu}) K_{\mu}(2\sqrt{J\mu})
\end{equation}
so that the resolution of unity is satisfied
\begin{equation}
\int|J,\gamma><J,\gamma|d\mu(J,\gamma) = 1
\end{equation}
3.Temporal stability:
\begin{equation}
e^{-\frac{iH^{1}t}{\hbar}}|J,\gamma> = |J,\gamma+\alpha t>
\end{equation}
4.Action identity:
\begin{equation}
<J,\gamma|H^{1}|J,\gamma> = \hbar \alpha J
\end{equation}
The overlapping of two coherent states is
\begin{equation}
<J^{\prime},\gamma'|J,\gamma> = \frac{\Gamma(\mu+1)}{N(J)N(J')}\sum_{n=0}^{\infty}\frac{(JJ'\mu^2)^{\frac{n}{2}}e^{i(\gamma'-\gamma)e_n^1}}{n!\Gamma(n+1+\mu)}
\end{equation}
If $\gamma = \gamma^{\prime}$, the overlapping is reduced to
\begin{equation}
<J^{\prime},\gamma|J,\gamma> = \frac{1}{\sqrt{I_{\mu}(2\sqrt{J\mu})I_{\mu}(2\sqrt{J'\mu})}}I_{\mu}(2(JJ'\mu^2)^{\frac{1}{4}})
\end{equation}
\section{\bf Revival dynamics}
In this section we shall study the revival dynamics of the coherent state (10). To demonstrate the role of the mass parameter $\lambda$ on the revival dynamics, all the figures below are drawn for a fixed value of $J =10$ and two different values of $\mu$ which is inversely proportional ($\mu = \frac{2}{|\Lambda|}$) to the mass parameter $\lambda$.
For a general wave packet of the form
\begin{equation}
|\psi(t)> = \sum_{n \geq 0}c_n e^{-iE_n t/\hbar}|n>
\end{equation}
with $\sum_{n \geq 0}|c_n|^2 = 1$, the concept of revival arises from the weighting probabilities $|c_n|^2$. For the coherent state (10), the weighting distribution is given by
\begin{equation}
|c_n|^2 = \frac{(J\mu)^{n+\frac{\mu}{2}}}{n!\Gamma(n+1+\mu)I_{\mu}(2\sqrt{J\mu})}
\end{equation}
Since the weighting distribution $|c_n|^2$ is crucial for
understanding the temporal behavior of the coherent state (10), we
show the curves of $|c_n|^2$ for $J=10$ and different $\mu$ in
Figure 1.

\begin{figure}[h]
\epsfxsize=3.2 in \epsfysize=2.5 in
\centerline{\epsfbox{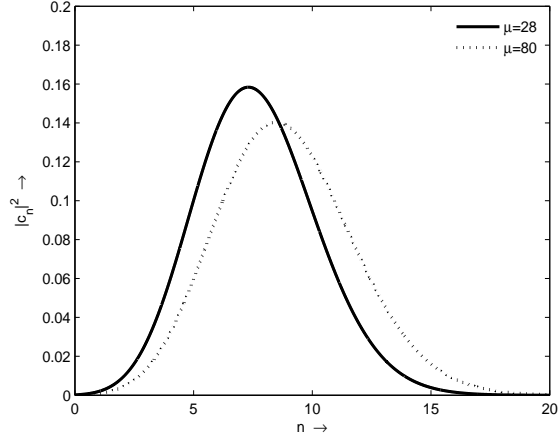}} \caption{ Plot of the weighted
distribution $|c_n|^2$ given in (26) for J=10 and $\mu=28,80.$}
\end{figure}
The Mandel parameter $Q$ is defined by
\begin{equation}
Q = \frac{(\Delta n)^2}{<n>} - 1
\end{equation}
where
\begin{equation}
\begin{array}{lcl}
<n> &=& \displaystyle \sum_{n=0}^{\infty} \frac{n J^n}{N(J)^2\rho_n}\\
\Delta n &=& [<n^2>-<n>^2]^{\frac{1}{2}}\\
\end{array}
\end{equation}
The Mandel parameter determines the nature of the weighting distribution function $|c_n|^2$. The case of $Q = 0$ coincides with the Poissonian weighting distribution $\displaystyle \frac{<n>^n e^{-<n>}}{n!}$ while the cases of $Q>0$ and $Q<0$ correspond to the super-Poissonian or sub-Poissonian statistics respectively. For the coherent state (10),
\begin{equation}
Q = \sqrt{J\mu}\left[\frac{I_{\mu+2}(2\sqrt{J\mu})}{I_{\mu+1}(2\sqrt{J\mu})} - \frac{I_{\mu+1}(2\sqrt{J\mu})}{I_{\mu}(2\sqrt{J\mu})}\right]
\end{equation}
where
\begin{equation}
<n> = \sqrt{J\mu} \frac{I_{\mu+1}(2\sqrt{J\mu})}{I_{\mu}(2\sqrt{J\mu})}
\end{equation}
In Figure 2 we plot the Mandel parameter $Q$ for $J=10$ and $\mu =
28,80$. It is evident from the figure that the Mandel parameter is
sub-Poissonian and it has been observed that it remains so for all
values $\mu$.

\begin{figure}[h]
\epsfxsize=3.2 in \epsfysize=2.5 in
\centerline{\epsfbox{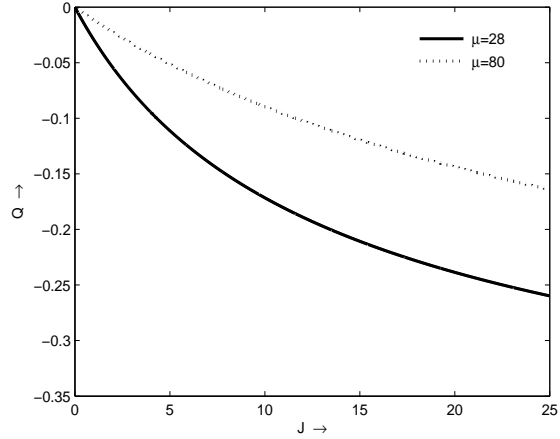}} \label*{}\caption{ Plot of the
Mandel parameter $Q$ given in (29) for$ J=10$ and $\mu=28,80.$}
\end{figure}
Now assuming that the spread $\Delta n = [<n^2>-<n>^2]^{\frac{1}{2}}$ is small compared to $<n> \approx \bar {n}$, we expand the energy $E_n^1$ in a Taylor series in $n$ around the centrally excited value $\bar {n}$:
\begin{equation}
E_n^1 \approx E_{\bar {n}}^1 + E_{\bar {n}}^{1'}(n-\bar {n}) + \frac{1}{2}E_{\bar {n}}^{1''}(n-\bar {n})^2 + \frac{1}{6}E_{\bar {n}}^{1'''}(n-\bar {n})^3 + \cdots
\end{equation}
where each prime on $E_{\bar{n}}^1$ denotes a derivative with respect to $n$. These derivatives define distinct time scales, namely the classical period $T_{cl} = \displaystyle \frac{2\pi\hbar}{|E_{\bar{n}}^{1'}|}$, the revival time $t_{rev} = \displaystyle \frac{2\pi\hbar}{\frac{1}{2}|E_{\bar{n}}^{1''}|}$and so on. For $E_n^1$ in Eqn.(9), $T_{cl} = \displaystyle \frac{2\pi\hbar}{\alpha(2\bar {n}+\mu)}$ and $t_{rev} = \displaystyle \frac{2\pi\mu}{\alpha}$. There is no superrevival time here because the energy is a quadratic function in $n$. It is convenient to describe the wavepacket dynamics by an autocorrelation function
\begin{equation}
A(t) = <\psi(x,0)|\psi(x,t)> = \sum_{n \geq 0}|c_n|^2e^{\frac{-iE_n^1t}{\hbar}}
\end{equation}
For the coherent state (10) the autocorrelation function (32) is given by
\begin{equation}
A(t) = <J,0|J,\alpha t> = \frac{\Gamma (1+ \mu)}{N(J)^2}\sum_{n \geq 0}\frac{(J\mu)^n}{n!\Gamma(n+1+\mu)}e^{-i\alpha(n+\frac{n^2}{\mu})t}~~
\end{equation}
From (33) it follows that $|A(t-t_{rev})|^2 = |A(t)|^2$ so that
the autocorrelation function is symmetric about
$\frac{t_{rev}}{2}$. In other words whatever happens in
$[0,\frac{t_{rev}}{2}]$ is repeated subsequently. We note that $0
\leq |A(t)|^2 \leq 1$ and it denotes the overlap between the
coherent state at $t=0$ and the one at time $t$. So, the larger
the value of $|A(t)|^2$, the greater the coherent state at time
$t$ will resemble the initial one. In Figure 3(a), 3(b) and 3(c)
we plot the squared modulus of the autocorrelation function (33)
in units of $t_{rev}$ for $J=10$ and $\mu = 1,28,80$ respectively.
Figure 3(a) clearly shows the full revival. The sharp peaks in
Figure 3(b) and 3(c) arise due to fractional revivals.

\begin{figure}[h]
\epsfxsize=6 in \epsfysize=4.5 in \centerline{\epsfbox{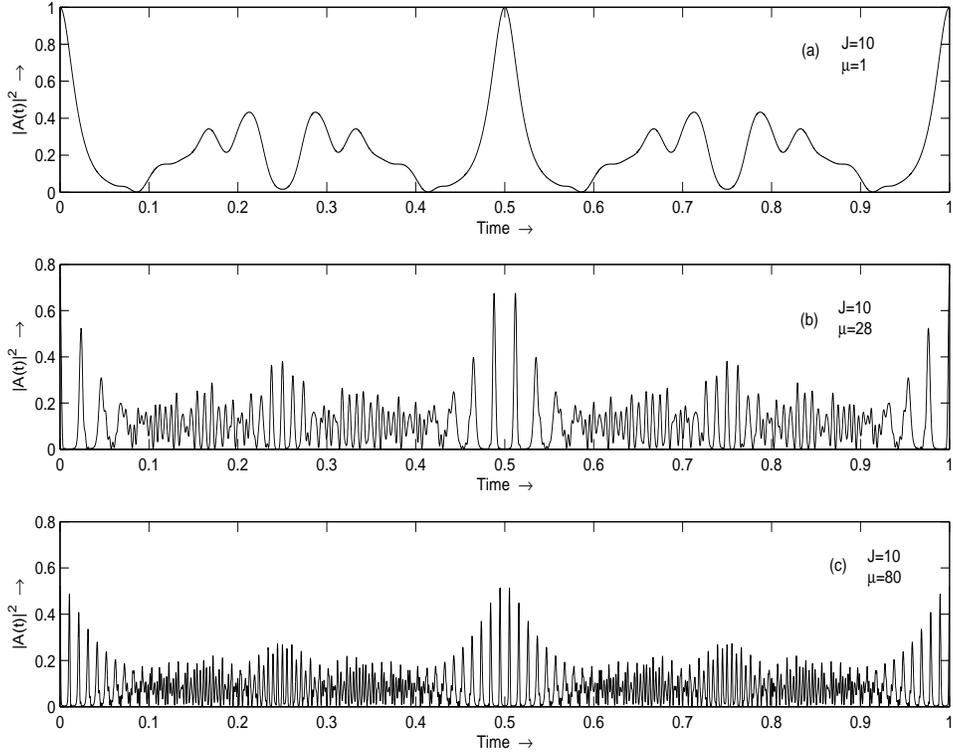}}
\label*{}\caption{ Plot of the square modulus of the
autocorrelation function $|A(t)|^2$ given in (33) for $(a) J=10,
\mu=1, ~(b) J=10,\mu=28, (c) J=10, \mu=80.$}
\end{figure}

\noindent
We can see from figures 3(a) to 3(c) that different values of $\mu$ lead to qualitatively different types of motion of the coherent wavepacket.\\
Now we shall study the mechanism of the fractional revival by phase analysis [86,87]. For this,we use the time scale determined by the period of the complete revival $t_{rev} = \frac{2\pi\mu}{\alpha} = 1$. So the phase of the $n$-th stationary state is
\begin{equation}
\phi_n(t) = 2\pi(\mu n + n^2)t
\end{equation}
where $\mu = \frac{2}{|\lambda|}$ is assumed to be an integer.\\
At arbitrary moments of time, the phases (34) of individual components of the packet are uniformly mixed so that it is not possible to make any definite conclusion about the value of the autocorrelation function. However, at specific moments, the distribution of phases can gain some order; for example, the phases can split into several groups of nearly constant values. The fractional revival of $q$-th order is defined as the time interval during which the phases are distributed among $q$ groups of nearly constant values.\\
To consider fractional revival of order $q$, in the vicinity of time $t = \frac{1}{q}$, it is convenient to write $n$ as $n = kq + \Delta$, where $k = 0,1,2,\cdots$ and $\Delta = 0,1,2,\cdots q-1$. Then the autocorrelation function can be written as
\begin{equation}
A(t) = \sum_{\Delta = 0}^{q-1}P_{\Delta}(t)
\end{equation}
and
\begin{equation}
P_{\Delta}(t) = \sum_k c_{kq+\Delta}e^{-i\phi_{kq+\Delta}(t)}
\end{equation}
where
\begin{equation}
c_{kq+\Delta}=\frac{(J\mu)^{kq+\Delta+\frac{\mu}{2}}}{(kq+\Delta)!\Gamma(kq+\Delta+1+\mu)I_{\mu}(2\sqrt{J\mu})}
\end{equation}
and $\phi_{kq+\Delta}(t)$ is given by (34) replacing $n$ by $kq+\Delta$. For quantum numbers that are multiples of the revival order i.e.$n = kq$, we have
\begin{equation}
\begin{array}{lcl}
\phi_{kq}(\frac{1}{q}) &=& 2\pi(k\mu + k^2q) = 0 (mod 2\pi)\\
P_0(\frac{1}{q}) &=& \sum_k c_{kq}
\end{array}
\end{equation}
When $\Delta \neq 0$, the phases of the corresponding states are
\begin{equation}
\begin{array}{lcl}
\phi_{kq+\Delta}(\frac{1}{q}) &=& 2\pi(\mu\Delta+\Delta^2)/q\\
P_{\Delta}(\frac{1}{q}) &=& exp[-2\pi i(\mu\Delta q^{-1}+\Delta^2q^{-1})\sum_k c_{kq+\Delta}
\end{array}
\end{equation}
where in obtaining (37) and (38) we have made use of the result
\begin{equation}
2\pi(k(\mu+2\Delta)+k^2 q) = 0 (mod 2\pi)
\end{equation}
Thus from (38) and (39) it follows that, around time
$t=\frac{1}{q}$ the Gazeau-Klauder coherent state (10) splits into
$q$ packet fractions such that the $\Delta = 0$ packet fraction
has zero phase while relative to this the other packet fractions
have a constant phase $(\mu \Delta q^{-1} + q^{-1}{\Delta}^2)$. It
must be mentioned that the way the packet is divided into
fractions is determined only by the order of the revival analyzed.
Figure 4 and Figure 5 give the time dependence of the
survival functions $|P_{\Delta}(t)|^2$ in units $t_{rev}$ for
$\Delta = 0,1,2,3$ during the whole period of the complete revival
for fixed $J=10$ and $\mu = 28,80$ respectively. It is to be noted
that fractions with odd and even $\Delta$ differ by their revival
periods.

\newpage
\begin{figure}[h]
\epsfxsize=8.5 in \epsfysize=4.5 in
\centerline{\epsfbox{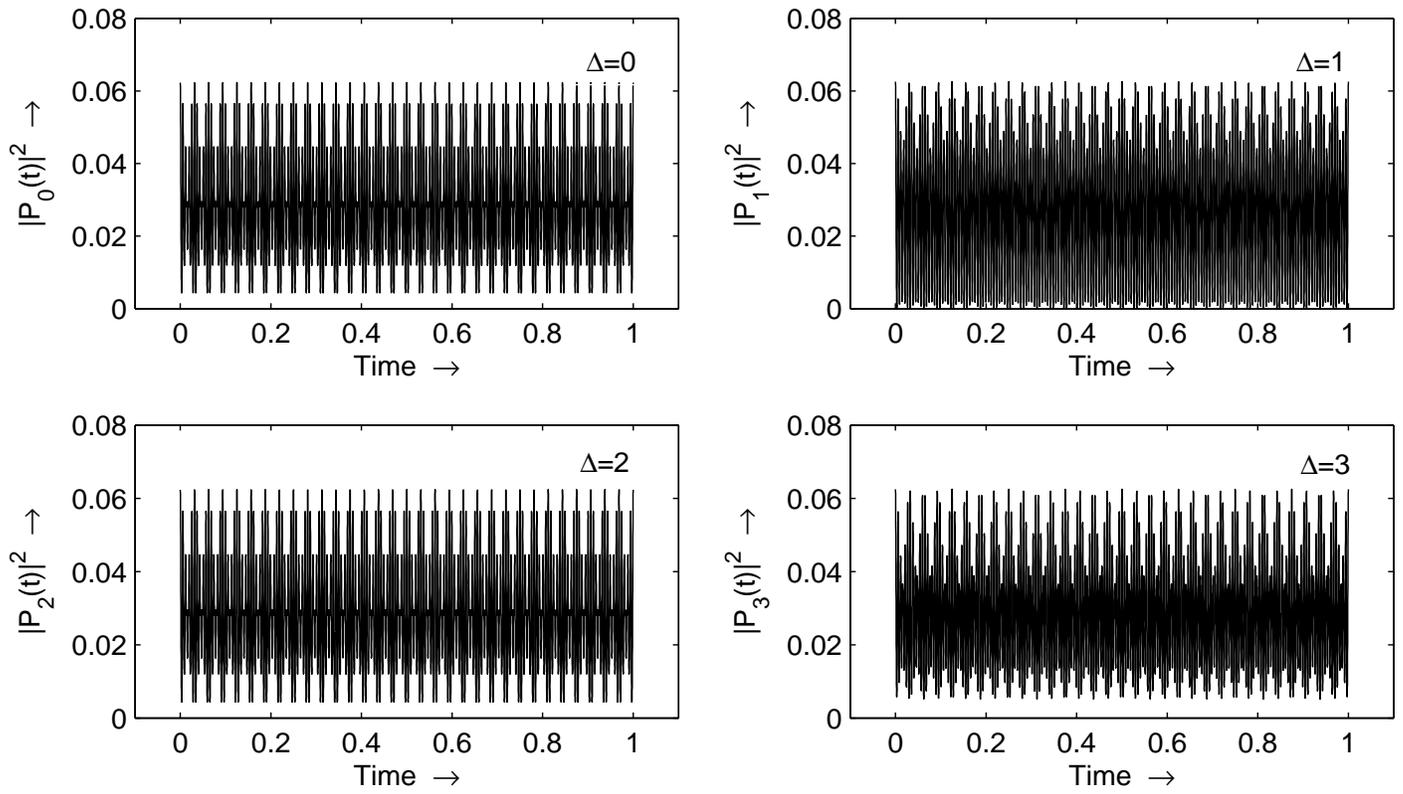}} \label*{}\caption{ Plot of the
Survival functions $|P_{\Delta}(t)|^2$ given in (37) for
$\Delta=0,1,2,3$ and $J=10, \mu=28 .$}
\end{figure}

\newpage
\begin{figure}[h]
\epsfxsize=8.5 in \epsfysize=4.5 in
\centerline{\epsfbox{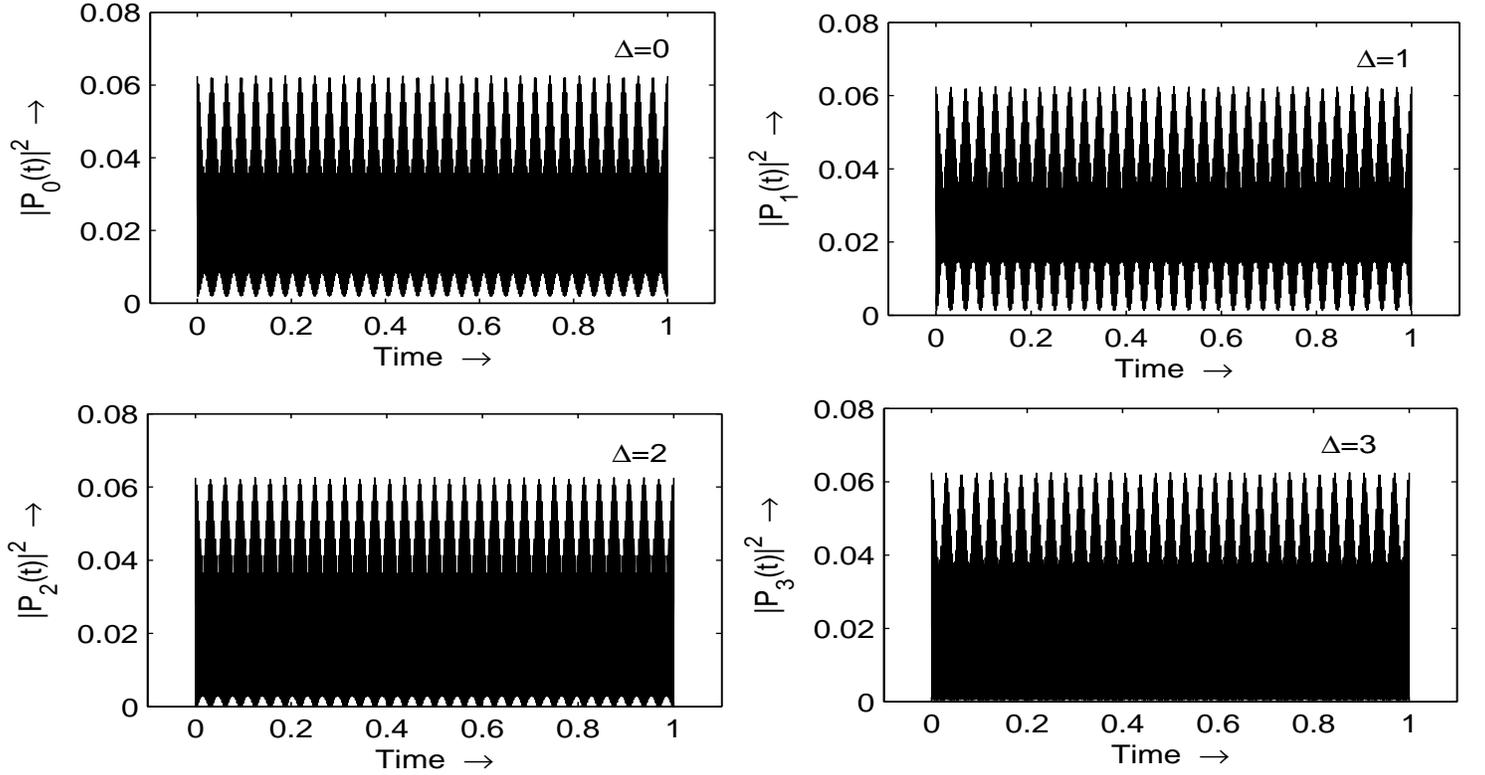}}\label*{} \caption{ Plot of the
Survival functions $|P_{\Delta}(t)|^2$ given in (37) for
$\Delta=0,1,2,3$ and $J=10, \mu=80.$}
\end{figure}

The intensity of the fractional revival can be defined as the
value of the wave packet survival function at the moment of
revival. This value is composed of two terms: the sum of survival
functions for the packet fractions ( the diagonal term) and the
interference term describing the interaction of the packet
fractions
\begin{equation}
S\left(\frac{1}{q}\right) = \sum_{\Delta = 0}^{q-1} |P_{\Delta}\left(\frac{1}{q}\right)|^2 + \sum_{\Delta = 0}^{q-1}\sum_{\Gamma \neq \Delta}P_{\Delta}\left(\frac{1}{q}\right)P_{\Gamma}^*\left(\frac{1}{q}\right)
\end{equation}
In Figure 6(a) and Figure 6(b) we present the time dependence of the diagonal terms in units of $t_{rev}$ for fixed $J=10$ and $\mu = 28,80$ respectively. It is seen from the figures that the diagonal term related to the individual packet fractions takes positive values only near the moments of fractional revival.\\
Figure 7(a) and Figure 7(b) show the time dependence of the
interference terms in units of $t_{rev}$ for fixed $J=10$ and $\mu
= 28,80$ respectively. It is seen from the figures that the
interference term plays a constructive, destructive or indifferent
role near the moments of fractional revival.
\newpage

\begin{figure}[h]
\epsfxsize=6 in \epsfysize=2.5 in
\centerline{\epsfbox{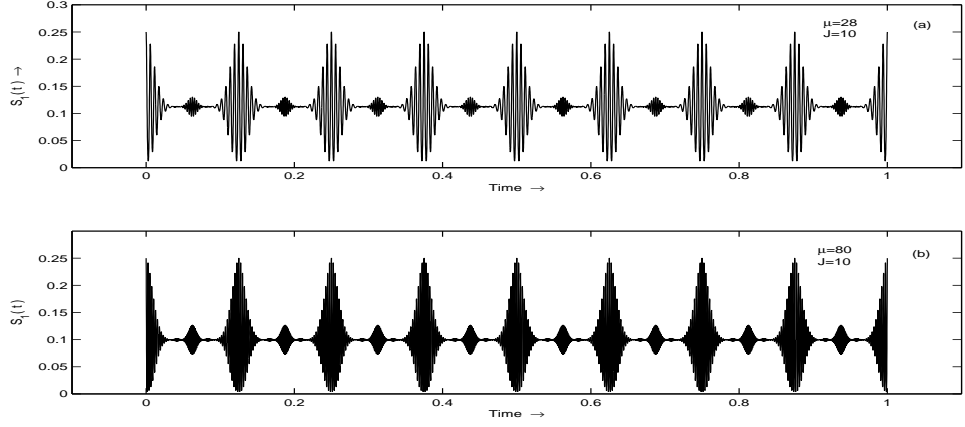}}\label*{} \caption{ Plots of the
diagonal term $S_1(t)=\Sigma_{\Delta=0}^3 |P_{\Delta}(t)|^2$ of
the survival function for $(a)J=10 , \mu = 28$ and $(b) J=10 ,
\mu=80.$}
\end{figure}

\begin{figure}[h]
\epsfxsize=6 in \epsfysize=2.5 in
\centerline{\epsfbox{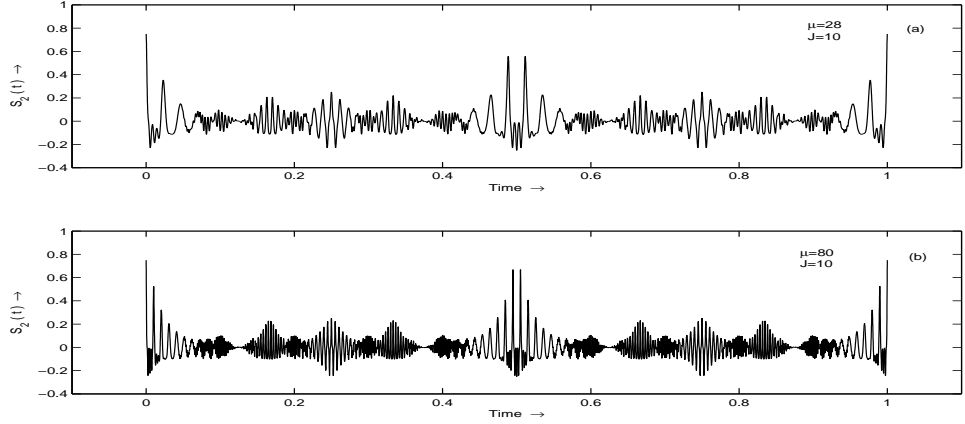}}\label*{} \caption{ Plots of the
interference term $S_2(t)=\Sigma_{\Delta=0}^3 \Sigma_{\Gamma\neq
\Delta}^{3} P_{\Delta}(t)P_{\Gamma}^\ast(t)$ of the survival
function.(a) $J=10 , \mu=28$ ; (b) $J=10 , \mu=80.$}
\end{figure}
\newpage
\section{\bf Summary}
In this paper we have constructed the coherent states for nonlinear oscillator via Gazeau-Klauder formalism [45,46].
 These coherent states are shown to satisfy the requirements of continuity of labeling, resolution of unity,
 temporal stability and action identity. The plots of the weighting distribution for these coherent states are almost
  Gaussian in nature. The Mandel parameter $Q$ is sub-Poissonian which indicates that the coherent state (10) exhibits
  squeezing for all values of $\mu$. The fractional revivals of the coherent states are evident from the Figure 3(b)
  and 3(c) of the squared modulus of the autocorrelation function. This is in contrast to the result obtained in ref [84] where the wave packet revival in an infinite well for the Schr\"odinger equation with position dependent mass [84] was studied. In ref [84] it was found that though full revival takes place, there is no fractional revival in the usual sense. Instead, a very narrow wave packet is located near one wall of the well, when the mass is higher. In the present paper the times of appearance, the spectral compositions and the intensities of the fractional revivals are determined by phase analysis.

\end{document}